\documentclass[]{fairmeta}

\usepackage{amssymb,amsthm,amsmath}
\usepackage{wrapfig}
\usepackage{mathtools}

 \usepackage{xspace}

\renewcommand{\phi}{\varphi}

\usepackage{algorithm}
\usepackage{algorithmic}
\usepackage{float}
\usepackage{listings} 
\usepackage{pmboxdraw} 
\usepackage{amssymb}   
\usepackage[T1]{fontenc}
\usepackage{lmodern}
\usepackage{graphicx}
\graphicspath{{assets/}}

\title{Sema Code: Decoupling AI Coding Agents into Programmable, Embeddable Infrastructure}

\author{Huacan Wang, Jie Zhou\textsuperscript{*}, Ningyan Zhu, Shuo Zhang, Feiyu Chen, Jiarou Wu, Ge Chen, Chen Liu, Wangyi Chen, Xiaofeng Mou, Yi Xu\textsuperscript{$\dagger$}}

\affiliation{Midea AIRC}

\abstract{
AI coding agents have become central to developer workflows, yet every existing solution locks its reasoning capabilities within a specific delivery form, such as a CLI, IDE plugin, or web application. This limitation creates systemic barriers when enterprises attempt to reuse these capabilities across heterogeneous engineering environments. To address this challenge, we present Sema Code, an open AI coding framework built on the principle of being embeddable, pluggable, and framework-first. Sema Code completely decouples the core agent engine from all client layers, publishing it as a standalone npm library that any runtime can drive programmatically. Built around this architecture, we designed eight key mechanisms: multi-tenant engine isolation, FIFO input queuing with safe session reconstruction, adaptive context compression, multi-agent collaborative scheduling, intelligent Todo-based process management, four-layer asynchronous permission control, three-tier ecosystem integration spanning MCP, Skills, and Plugins, and a background task framework with separated execution and observation privileges. These mechanisms collectively address the engineering challenges of transforming a complex agent engine into a shared, programmable core. Demonstrating its architectural versatility, the same Sema Core engine simultaneously powers a VSCode extension and a multi-channel messaging gateway, which we name SemaClaw, to unify agent interactions across platforms such as Telegram and Feishu. These represent two fundamentally different product forms sharing an identical reasoning kernel, differing only at the client layer.

}

\date{
    April 10, 2026 \\[0.5em]
    \begin{tabular}{@{}l l@{}}
        \textbf{GitHub:} &
        \href{https://github.com/midea-ai/sema-code-core}{https://github.com/midea-ai/sema-code-core} \\
        &
        \href{https://github.com/midea-ai/sema-code-vscode-extension}{https://github.com/midea-ai/sema-code-vscode-extension}
    \end{tabular}
    \\[0.6em]
    \mbox{}
}

\begin{document}

\begin{tikzpicture}[remember picture, overlay]
\node[anchor=south west, xshift=2cm, yshift=0.5cm, align=left, ] at (current page.south west) {%
    \quad
    \textsuperscript{*} Primary developer. 
    \textsuperscript{\textdagger} Corresponding author.%
};

\end{tikzpicture}

\maketitle

\begin{tikzpicture}[remember picture, overlay]
\node[anchor=north east, xshift=-3.5cm, yshift=-13.2cm] at (current page.north east) {%
    \includegraphics[height=4.5em]{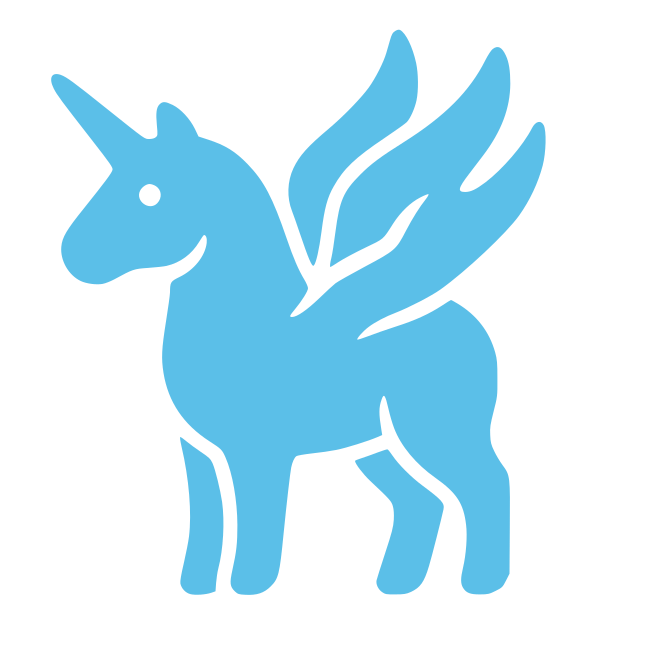}%
};
\end{tikzpicture}

\section{Introduction}

Over the past two years, AI coding agents have leapt from laboratory prototypes to commercial products. Since OpenAI Codex~\cite{chen2021codex} first demonstrated the potential of LLMs for code generation, the technology has advanced so rapidly that autonomous systems such as SWE-agent~\cite{yang2024sweagent} and Agentless~\cite{xia2024agentless} can now resolve real-world GitHub issues~\cite{yang2024swebench}. Simultaneously, mainstream developer tools have experienced widespread adoption. GitHub Copilot~\cite{peng2023copilot_impact} has surpassed one million monthly active developers, with controlled experiments showing a 55.8\% improvement in task completion speed. Cursor~\cite{cursor2024} has redefined how editors interact with AI, while Claude Code~\cite{anthropic2025claude_code} has brought agentic coding to the terminal. A consensus has formed: AI coding is no longer an optional enhancement but a core multiplier of engineering productivity.

\begin{figure}[t]           
    \centering              
    \includegraphics[width=0.9\columnwidth]{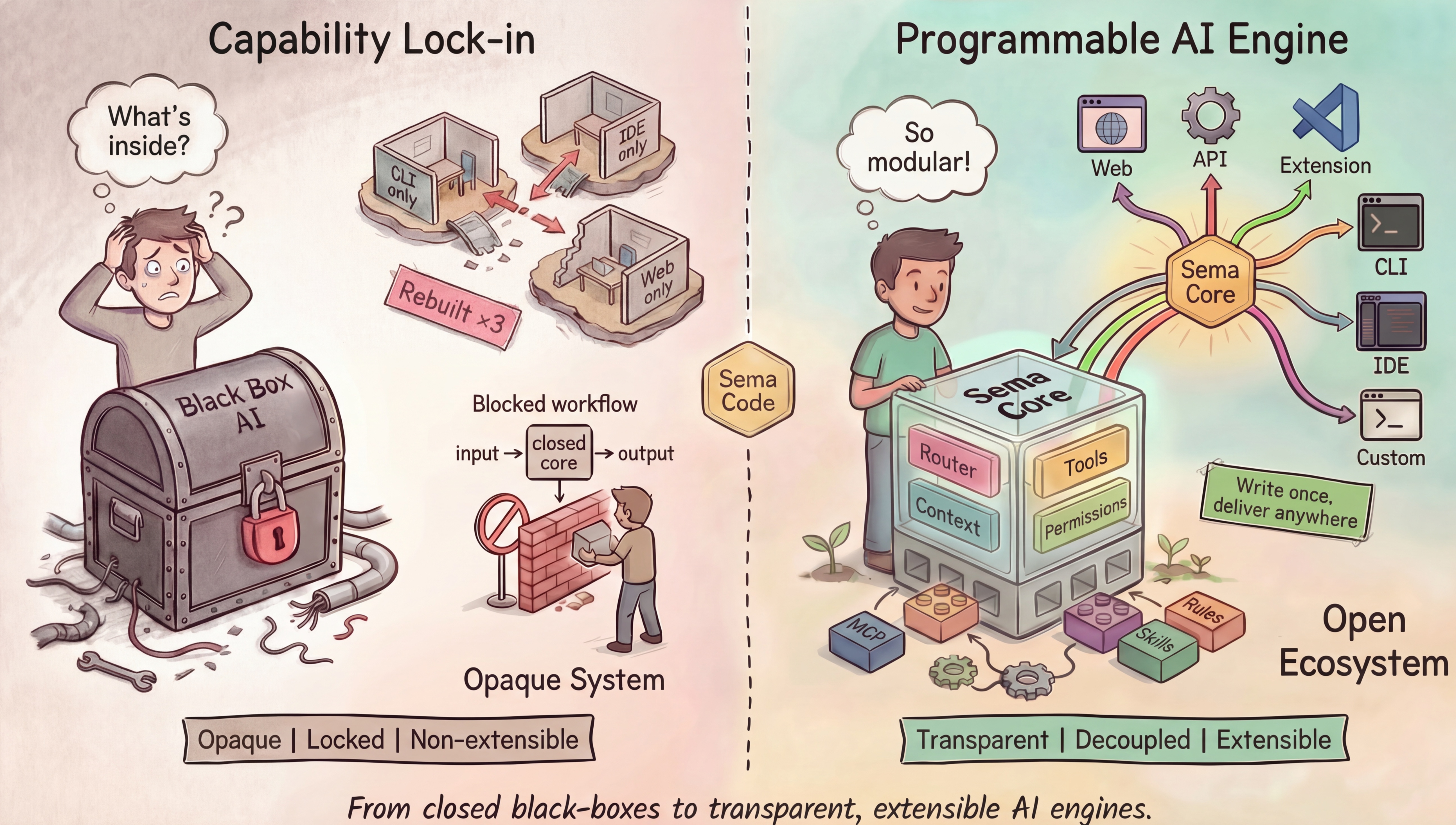}
    \caption{Architectural comparison between traditional, monolithic AI systems and the proposed Sema Code framework, illustrating the shift from opaque capability lock-in to a transparent, decoupled, and extensible AI engine.}
    \label{fig:background}
\end{figure}

Yet, when enterprises attempt to integrate these agents into their own infrastructure, a structural contradiction surfaces. Every mainstream AI coding solution deeply couples its frontend interaction with its core reasoning mechanics: Claude Code's agentic loop ships monolithically with its CLI layer, Cursor's intelligent engine is welded to a custom VSCode fork, and GitHub Copilot's functional boundaries are dictated by its editor plugin form factor. As a result, enterprises face significant barriers when attempting to embed these advanced features, such as autonomous code generation, multi-step reasoning, and contextual understanding, into backend services. Similarly, they struggle to replace the underlying model for private deployment or to serve multiple channels, such as IDEs, web apps, and messaging bots, from a single intelligent engine, as illustrated in Figure~\ref{fig:background}. This occurs because the core reasoning logic remains fundamentally inseparable from the specific product forms that deliver it.



Existing solutions package AI coding capabilities as \textit{an application to be used}, not \textit{an engine to be embedded}. Applications are optimized for a specific user experience and inherently resist decomposition; engines prioritize programmability, allowing any host system to consume their capabilities. SQLite~\cite{gaffney2022sqlite} is embedded in billions of devices; Chromium powers everything from browsers to Electron apps. Both succeeded precisely because they were designed as engines, not applications. \textbf{Can the core reasoning engine of AI coding follow the same path---from being encapsulated within applications to being programmatically driven by any engineering system?}

To this end, we propose \textbf{Sema Code}, a lightweight pluggable AI coding framework. Its core insight is straightforward: decouple the agent engine from all client layers, publish it as a standalone library, and make ``driving an AI coding agent'' as simple as ``connecting to a database.'' Around this goal, we systematically address the engineering challenges that arise when delivering a core engine as a programmable interface: multi-tenant isolation, context lifecycle management, multi-agent coordination, permission security, and ecosystem compatibility. To validate this architecture, we deployed the Sema Core engine to simultaneously power a VSCode extension and a multi-channel agent platform (Telegram, Feishu, etc.), sharing an identical reasoning kernel with differences only at the client layer.

In summary, our core contributions are as follows:

\begin{itemize}
\item We propose an embeddable engine architecture that decouples the system into client, core engine, and service layers. This allows the complete AI coding agent to be packaged as a UI-free \texttt{npm} library, supporting direct imports, WebSocket, and cross-language gRPC integration.
\item We design three core engine mechanisms to ensure state safety and resource efficiency under concurrent workloads: multi-tenant state isolation, FIFO input queuing with session reconstruction, and adaptive context compression.
\item We introduce an agent runtime comprising multi-agent collaborative scheduling with isolated sub-agent state spaces and shared interrupt control, ID-matched "Todo" process management, and a background task system that decouples execution privileges from observation rights, forming a complete task execution and scheduling layer.
\item We present a four-layer asynchronous permission system that balances security compliance with operational fluency, complemented by an extensible three-tier ecosystem (supporting MCP, Skill, and Plugin standards) and a multi-model adaptation layer for seamless capability integration.
\end{itemize}


\section{Related Work}

\textbf{Foundational capabilities.} The core capabilities of modern AI coding systems rest on a series of foundational advances. Chain-of-Thought~\cite{wei2022cot} revealed that intermediate reasoning steps unlock complex task performance in LLMs; ReAct~\cite{yao2023react} alternates reasoning and action, enabling simultaneous chain-of-thought and environment interaction---now the core paradigm of modern agent systems. In tool use, Toolformer~\cite{schick2023toolformer} demonstrated self-supervised tool invocation learning; Gorilla~\cite{patil2024gorilla} and ToolLLM~\cite{qin2024toolllm} scaled tool invocation to thousands of real APIs. In context management, MemGPT~\cite{packer2023memgpt} adapted OS virtual memory paging for hierarchical context management of fixed-window LLMs; LLMLingua~\cite{jiang2023llmlingua,li2025prompt_compression_survey} explored token-level prompt compression. These studies provide critical technical foundations for AI coding framework design while revealing a common gap: most work targets algorithmic capabilities rather than system-level production challenges. Recent surveys on LLM agents~\cite{wang2024agent_survey} and LLM for software engineering~\cite{hou2024llm4se_survey} provide panoramic views of this rapidly evolving landscape.

\textbf{Developer-facing general frameworks.}
OpenHands~\cite{wang2024openhands} is a representative open-source multi-agent coding framework that performs well on SWE-bench~\cite{yang2024swebench}, supporting sandbox environments and custom tools. Its design, however, targets research scenarios, with production-grade concerns---multi-tenant isolation, permission management, cross-language integration---still under-addressed. MetaGPT~\cite{hong2024metagpt} models software engineering as multi-role collaboration, encoding human SOPs into structured agent communication protocols for end-to-end automation from requirements to code generation, though integration with existing IDE workflows remains a long path. ChatDev~\cite{qian2024chatdev} takes a similar approach, organizing multi-agent development through chat chain topologies. AutoGen~\cite{wu2023autogen} provides a flexible multi-agent conversation framework for composable LLM applications, but its general-purpose design lacks coding-specific optimizations. LangChain~\cite{chase2022langchain} offers general LLM orchestration with heavy abstraction layers, introducing debugging complexity and performance overhead in coding-specific scenarios.

\textbf{End-user product systems.}
Claude Code~\cite{anthropic2025claude_code} is among the most capable terminal-based AI coding assistants, featuring a complete agentic loop, rich tool invocation, and an open plugin ecosystem. Designed for individual developers seeking an optimal CLI experience, its deep integration of core logic with the CLI layer makes it difficult to embed in heterogeneous runtimes. Cursor~\cite{cursor2024} deeply customizes VSCode, tightly fusing code completion, Chat, and Composer with the editor. The user experience is polished, but the closed architecture limits embedding flexibility for third-party systems. GitHub Copilot~\cite{github2023copilot} leverages the GitHub ecosystem through code completion and inline chat, with a recently introduced agent mode. Its capability boundaries, however, are constrained by the editor plugin form factor, with limited support for enterprise private deployment and model replacement. In autonomous coding, SWE-agent~\cite{yang2024sweagent} introduced the Agent-Computer Interface (ACI), enabling LLM agents to navigate repositories, edit files, and run tests autonomously; Agentless~\cite{xia2024agentless} achieves competitive results through a concise locate-fix-verify pipeline without granting autonomous decision-making authority. These paradigms represent opposite extremes of agent autonomy, yet both lock capabilities within their respective execution environments. The common thread: polished user experiences, but core capabilities tightly bound to delivery form, making programmatic reuse by third-party systems difficult.

The divergence between these lines reveals an unfilled gap: an AI coding framework combining product-grade engineering quality with embeddability by any runtime. Sema Code targets this gap. Unlike product systems, it publishes the core engine as a standalone library without presupposing delivery form; unlike general frameworks, it provides full-stack engineering for AI coding---from multi-tenant isolation to permission management, from context lifecycle to ecosystem integration---as a production-ready solution.


\begin{figure}[t]
\centering
\includegraphics[width=0.8\linewidth]{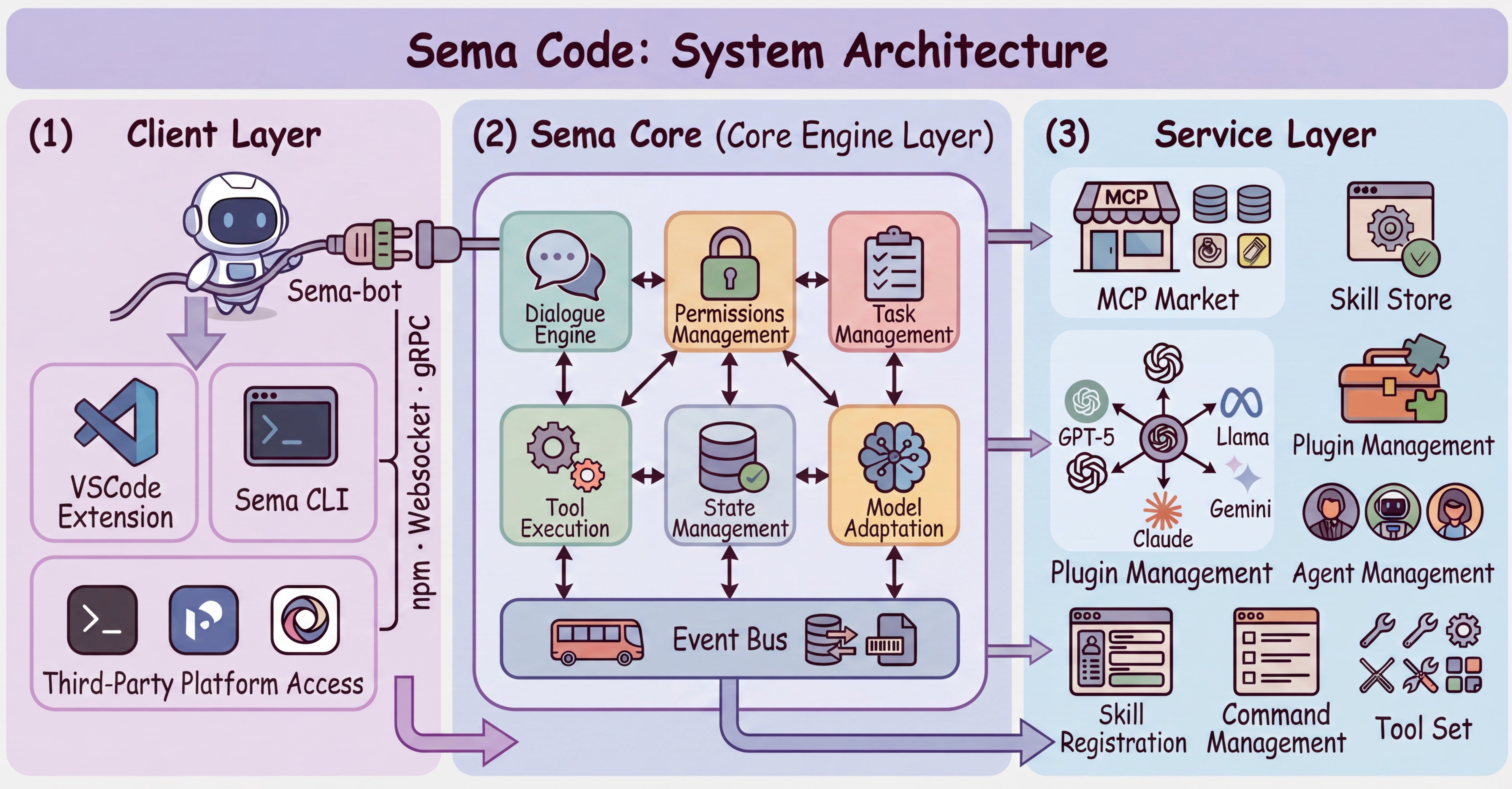}
\caption{Sema Code three-layer architecture. The core engine layer contains no UI code and ships as a standalone npm package. The client layer interacts with the engine exclusively through event-driven public APIs, enabling the same engine to power fundamentally different delivery forms.}
\label{fig:architecture}
\end{figure}


\section{System Design}

Turning an AI coding agent into an embeddable engine introduces two categories of challenges that do not arise in single-user, single-product settings. First, when multiple tenants share a single engine process, their state---including conversation history, tool permissions, and abort signals---must be strictly isolated without incurring the overhead of separate OS processes. Second, long-running agentic tasks operate under finite context windows, so the engine must manage the context lifecycle autonomously. Section~\ref{sec:arch} presents the three-layer separation architecture that decouples the engine from all clients, and Section~\ref{sec:core_mechanisms} describes the core engine mechanisms that address these challenges.

\subsection{Three-Layer Separation Architecture}
\label{sec:arch}

Sema Code follows a \textbf{three-layer separation principle} consisting of a client layer, a core engine layer, and a service layer, as shown in Figure~\ref{fig:architecture}.

The central design decision is that Sema Core encapsulates all reasoning, tool invocation, and state management logic, while containing no UI code or runtime-specific assumptions. Clients handle their own rendering or channel forwarding by subscribing to the engine's event stream. We chose an event-driven architecture rather than an RPC-first design because agentic tasks produce heterogeneous and interleaved outputs---including text fragments, tool results, permission requests, and progress updates---that map naturally onto a typed event stream. By contrast, an RPC model would require callers to poll for updates or coordinate across multiple endpoints. This decoupling reduces the effort of adding a new form factor from replicating the entire application to simply subscribing to an event stream and implementing presentation logic. This design echoes SQLite's modular architecture, which separates the SQL compiler, core, and backend~\cite{gaffney2022sqlite}, as well as the Language Server Protocol's separation of language intelligence from IDEs~\cite{barros2022lsp}.

Sema Core is implemented in the JavaScript ecosystem. To enable integration with systems written in other languages, it exposes WebSocket and gRPC as cross-language interfaces. Clients first establish a session and receive an opaque session token, then invoke the corresponding engine instance through a streaming interface that returns heterogeneous events, including text chunks, tool results, token statistics, permission requests, and session completion markers. Through this interface, systems written in Java, C\#, Python, or other languages can access full agent capabilities without requiring changes to their existing architecture.



\subsection{Core Engine Mechanisms}
\label{sec:core_mechanisms}

Exposing the engine as an embeddable programmable interface requires addressing two foundational challenges: maintaining state isolation and concurrency safety in multi-tenant environments, and managing context lifecycles during long-running agentic tasks. To see how these challenges manifest in practice, consider a concrete deployment where a single Sema Core process simultaneously powers a VSCode extension for individual developers and a messaging bot for a 50-person team. In such a dense environment, the engine must safely orchestrate dozens of concurrent sessions—preventing one user's quick, one-shot query from corrupting another's multi-step coding task, while also ensuring that these complex, cross-file operations do not exhaust the model's context window.

To resolve the complexities of these concurrent workloads, we organize our core engine mechanisms around two primary pillars. First, to manage multi-tenant execution, we implement a multi-tenant isolation model to prevent cross-user state contamination, a layered state hierarchy to balance agent-local autonomy with session-level control, and a FIFO input queue to guarantee atomic session transitions during message bursts. Second, to sustain long-context execution, we introduce an adaptive context tracking mechanism and a dual-path degradation policy. Collectively, these mechanisms ensure that long-running sessions remain within the model’s operational boundaries without sacrificing reasoning coherence. We describe these mechanisms in detail in the following.



\subsubsection{Multi-Tenant Isolation} %
\label{sec:isolation}

When Sema Core is embedded in a multi-tenant environment, such as an agent platform serving multiple users simultaneously, multiple engine instances must run concurrently within a single process. Under a traditional global singleton architecture, $n$ agent instances $\mathcal{A} = \{A_1, A_2, \ldots, A_n\}$ share a single state space $\mathcal{S}$. Consequently, any write operation by $A_i$ ($A_i \xrightarrow{w} \mathcal{S}$) might be observed by $A_j$'s read ($A_j \xrightarrow{r} \mathcal{S}$, where $j \neq i$). This shared state risks unpredictable state contamination, such as conversation histories leaking across user contexts or cross-instance interrupt collisions.

To mitigate these risks, Sema Core implements per-instance isolation through \textbf{asynchronous context tracking}, leveraging the native \texttt{AsyncLocalStorage} (ALS) mechanism in Node.js. ALS provides a continuation-local storage that persists across asynchronous execution boundaries.

Each engine instance $E_i$ is bound to a dedicated resource bundle, encompassing its internal event bus, state manager, tool orchestrator, and tenant configurations. By utilizing ALS, all asynchronous function calls within the lifecycle of $E_i$ are automatically associated with this local bundle without requiring explicit parameter passing. ALS ensures the reliable propagation of these resource references across complex asynchronous control flows (e.g., parallel task executions and deferred callbacks), guaranteeing strict isolation between any two instances. To ensure robustness and backward compatibility, the engine employs a two-tier resolution strategy: it first attempts to retrieve resources from the active ALS context, reverting to a global singleton only if no local context is available.

While ALS successfully isolates resources at the engine-instance level, each instance still requires fine-grained internal state organization. This is especially critical in scenarios where sub-agents operate alongside the main agent within the same session, as discussed in the following subsection.

\subsubsection{Hierarchical State Partitioning}

While ALS isolates resources at the instance level (across different users), a single instance often orchestrates a main agent and multiple sub-agents working concurrently within the same session. To prevent these sub-agents from polluting each other's context while maintaining unified control, we design a dual-layered state hierarchy.

The complete state space of any session, denoted as $\mathcal{S}_{\text{session}}$, is strictly partitioned into two tiers:
\begin{equation}
\mathcal{S}_{\text{session}} = \langle \mathcal{S}_{\text{local}},\ \mathcal{S}_{\text{global}} \rangle
\end{equation}
\textbf{Agent-Local State ($\mathcal{S}_{\text{local}}$):} This tier maintains strict isolation among all participating agents. For any given agent $a_i$ (whether main or sub-agent), its private state is mathematically formalized as a tuple:\begin{equation}\mathcal{S}_{\text{local}}(a_i) = \langle e_i,\ H_i,\ T_i,\ F_i \rangle\end{equation}where $e_i \in \{\text{idle},\ \text{processing}\}$ tracks the real-time execution status, $H_i$ maintains the isolated conversation history, $T_i$ stores the agent-specific task list (Todos), and $F_i$ records localized file-read timestamps. This localized tier guarantees that a sub-agent's intermediate reasoning steps or verbose tool outputs never bleed into the main agent's context.

\textbf{Session-Global State ($\mathcal{S}_{\text{global}}$):} Conversely, this tier encapsulates shared control primitives required to coordinate the entire agent tree. It primarily consists of a global permission flag $g_{\text{edit}}$ (dictating system-wide file modification rights) and a centralized abort controller $\mathcal{C}_{\text{abort}}$. By elevating these specific controls to the global tier, the system ensures that a single user cancellation signal reliably propagates to and terminates all active sub-agents simultaneously.

Having established robust spatial isolation---both across distinct user instances and among internal sub-agents---the architectural focus must shift to \emph{temporal} challenges. In multi-channel deployments, users frequently issue follow-up commands or switch contexts while the engine is actively computing a prior request.



\subsubsection{FIFO Input Queue and Safe Session Reconstruction}
\label{sec:fifo}

In multi-channel deployments (e.g., collaborative messaging bots), user input is inherently asynchronous and unpredictable. Users frequently fire multiple follow-up messages while the engine is still computing a previous response, or abruptly switch to entirely new projects mid-execution. To handle these concurrent events without dropping messages or corrupting the local state ($\mathcal{S}_{\text{local}}$), the engine relies on a unified FIFO dispatch mechanism coupled with safe session reconstruction protocols.

All incoming queries $q$ are routed through a state-aware dispatch function. Depending on the engine's current execution status $S_{\text{state}}$, inputs are either processed immediately or buffered:
\begin{equation}
\text{dispatch}(q) = \begin{cases} \text{enqueue}(Q, q) & \text{if}\ S_{\text{state}} = \text{processing} \\ \text{startQuery}(q) & \text{if}\ S_{\text{state}} = \text{idle} \end{cases}
\end{equation}

\textbf{Semantic Batching Policy:}
Simply dequeuing messages one by one would result in fragmented context and redundant API calls. Instead, the engine employs a semantic batching strategy. During a dequeue operation, if the head of the queue is a standard text message, the engine eagerly extracts and merges all subsequent non-command messages into a single unified prompt. Conversely, system commands (e.g., those prefixed with \texttt{/}) are dequeued and executed strictly individually. This policy optimizes LLM interaction rounds while preserving the strict isolation of operational commands.

\textbf{Safe Session Reconstruction:}Handling mid-execution task switches poses a severe risk of state leakage. If a user creates a new session while the agent is actively generating code, forcibly killing the process might leave residual files or corrupted memory states. Instead, Sema Core employs a staging mechanism: the new session ID is temporarily stored as \texttt{pendingSession}, and an abort signal is sent to the active $\mathcal{C}_{\text{abort}}$ controller. The aborted task's \texttt{finally} block detects this pending flag, entirely purges the current $\mathcal{S}_{\text{local}}$ state, and seamlessly initializes the new session. This guarantees a clean, safe reconstruction with zero state residue, regardless of when the interruption occurred.

While the input queue ensures that rapid user requests are never lost, this buffering mechanism accelerates the accumulation of conversation history. In long-running tasks, this rapid growth inevitably exhausts the language model's maximum context window, necessitating the adaptive compression strategy discussed next.





\subsubsection{Adaptive Context Compression}
\label{sec:compression}

The context window of large language models is a strictly finite resource. While context extension techniques such as YaRN~\cite{peng2024yarn} have pushed native limits beyond 128K tokens, the unbounded accumulation of historical messages in long-running agentic tasks inevitably exhausts these bounds. Prior works like MemGPT~\cite{packer2023memgpt} address this via OS-inspired virtual memory paging. However, these approaches require the LLM to explicitly orchestrate page-in and page-out operations, introducing severe tool-call overhead and latency at every context boundary. Conversely, naive truncation strategies lead to catastrophic forgetting, causing agents to lose track of prior codebase modifications or repeat previous steps.

To manage context efficiently without the overhead of extra LLM calls, Sema Core leverages cumulative usage metadata (e.g., the \texttt{input\_tokens} field in Anthropic APIs) for zero-cost tracking. Rather than recalculating the token footprint of the entire message array at every turn, an operation that incurs $O(k)$ complexity, the engine simply extracts the cumulative token count embedded in the most recent assistant message. This approach natively achieves $O(1)$ monitoring.

To calculate the effective context size, the engine adds an empirical forward buffer of 8,000 tokens to this cumulative count. This buffer is precisely sized to absorb the footprint of the upcoming interaction, encompassing system prompts, user queries, and tool-call/response pairs, while leaving sufficient headroom for unusually large tool outputs.

With this real-time metering in place, the engine continuously evaluates the effective context size and triggers compression when it exceeds 75\% of the model's maximum context limit $L$. This 25\% safety margin ensures the LLM has sufficient remaining space to generate multi-step plans and execute tool calls before hitting hard limits, preventing the premature discard of useful history.

Upon triggering, the algorithm first identifies the most recent user-initiated message (excluding tool results) as the pivot $m_{\text{pivot}}$. It then partitions the history into a compressible segment $\mathcal{H}_{\text{hist}}$ and an intact current turn $\mathcal{H}_{\text{keep}}$, thereby strictly preserving the structural pairing of recent tool calls and results.

\textbf{Primary Path (Semantic Summarization):} The engine prompts the LLM to generate a structured summary of $\mathcal{H}_{\text{hist}}$. This process is modeled as an information preservation problem, aiming to retain critical entities (e.g., file paths, function signatures, modification history) while aggressively discarding verbose tool outputs and intermediate reasoning traces:
\begin{equation}
\mathcal{H}^* = \arg\max_{\mathcal{H}'} \mathcal{I}(\mathcal{H}';, \mathcal{H}_{\text{hist}}) \quad \text{s.t.} \quad |\mathcal{H}'|_{\text{token}} \ll |\mathcal{H}_{\text{hist}}|_{\text{token}}
\end{equation}Following summarization, the engine applies three deterministic state corrections: stripping internal thinking blocks, recalibrating the cumulative usage metric, and re-injecting core rules and Todo reminders to maintain contextual integrity for subsequent reasoning.

\textbf{Degradation Path (Safe Truncation):} If the summarization process fails or times out, the engine gracefully degrades to deterministic truncation. Instead of naive halving, the algorithm scans the usage metadata to find the earliest assistant message that reduces the remaining context footprint to exactly half of the maximum window. Truncating exclusively at assistant messages ensures that the API's strict role-alternation constraints are preserved, guaranteeing a safe recovery.

Sub-agents bypass this compression mechanism entirely; their localized context lifecycles are strictly orchestrated by the main agent to prevent cascading compression conflicts.

\section{Agent Runtime}

While the mechanisms described above guarantee multi-tenant safety and context stability, executing complex agentic workloads introduces a different set of architectural challenges. Large-scale software engineering tasks, such as cross-module refactoring, inherently demand concurrency, specialization, and long-term state management. A monolithic, single-threaded agent loop is ill-equipped to simultaneously decompose these tasks, track granular progress across dozens of tool invocations, and execute long-running builds without blocking the conversational interface.

To address these limitations, the Sema Core runtime introduces a hierarchical execution architecture. The multi-agent scheduling framework for dynamic task decomposition is detailed in Section~\ref{sec:multiagent} and Section~\ref{sec:interrupt}. To ensure that execution remains observable to the user, Section~\ref{sec:todo} introduces a structured process manager for deterministic progress tracking. Finally, Section~\ref{sec:bgtask} explains how the system safely offloads time-consuming operations through a dedicated background execution engine.

\subsection{Multi-Agent Collaborative Scheduling}
\label{sec:multiagent}

\begin{figure}[t]           
    \centering              
    \includegraphics[width=0.9\columnwidth]{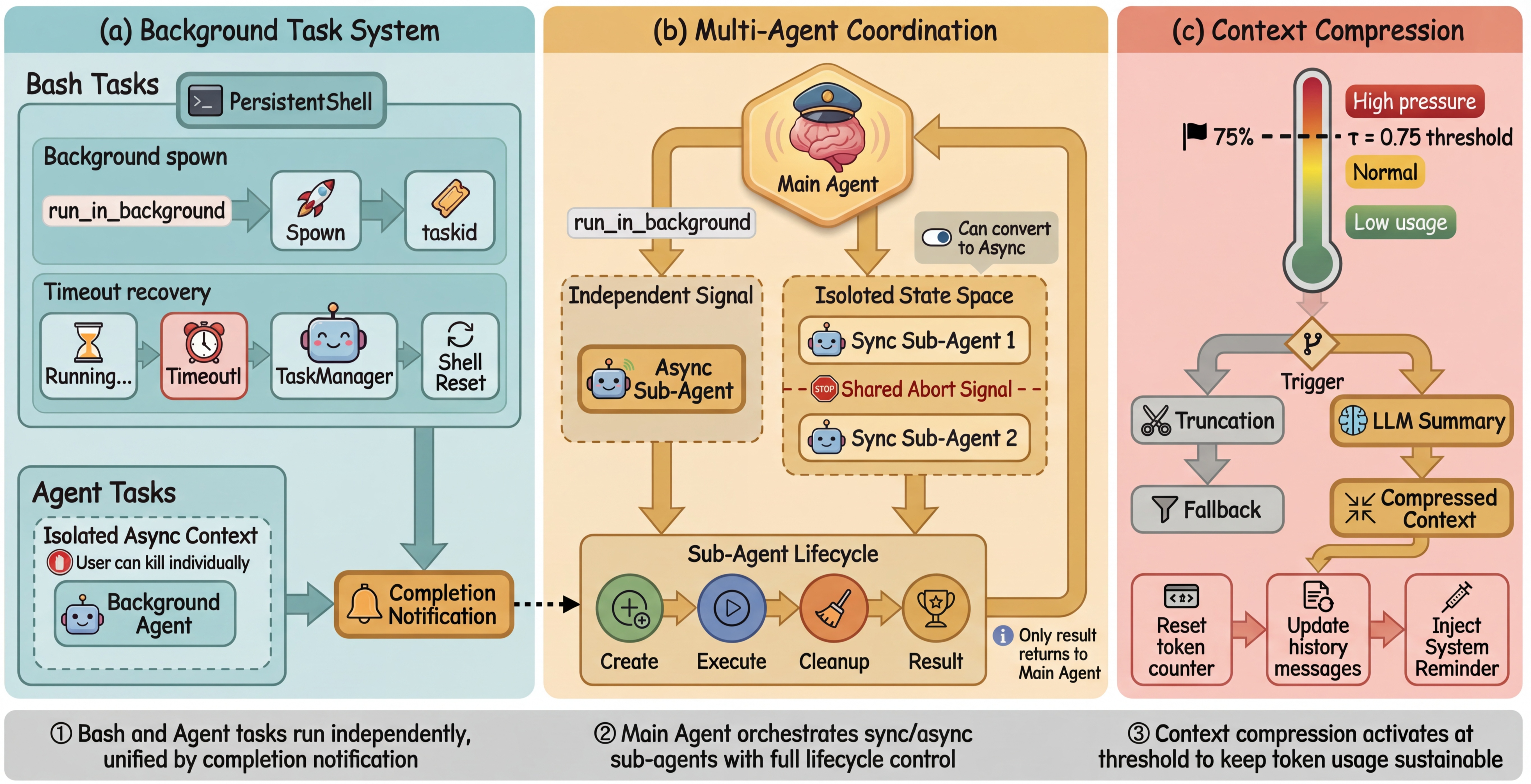}
    \caption{Overview of agent runtime, illustrating three core part: (a) the background task execution system, which asynchronously manages long-running Bash and Agent processes with timeout recovery; (b) the multi-agent coordination framework, featuring isolated state spaces, strict lifecycle management, and shared abort signals; and (c) the adaptive context compression module, which triggers dual-path token reduction upon reaching a high-pressure threshold.}
    \label{fig:runtime}
\end{figure}

Complex coding tasks decompose into subtasks with different capability requirements~\cite{hong2024metagpt,qian2024chatdev}. Processing all subtasks serially within a single agent context creates two concrete problems: first, Skill prompt instructions for one subtask---such as a code review checklist---contaminate the system prompt of subsequent subtasks like test generation, producing off-target outputs; second, subtask histories accumulate in the main context, accelerating compression triggers and degrading reasoning quality.

Existing multi-agent frameworks address task decomposition differently. MetaGPT~\cite{hong2024metagpt} encodes human software engineering SOPs into fixed role assignments (product manager, architect, engineer), coupling agents to predetermined workflows. ChatDev~\cite{qian2024chatdev} organizes collaboration through chat chain topologies where agents communicate via structured dialogue phases. Both approaches impose rigid coordination structures suited to end-to-end software generation but poorly matched to the ad-hoc, developer-initiated tasks typical of interactive coding assistants. Sema Code takes a different approach: isolated state spaces with shared interrupt control. Sub-agents share no conversation history or tool results with the main agent---only the abort controller $\mathcal{C}_{\text{abort}}$---allowing the main agent to delegate freely without predefined role schemas while retaining the ability to halt the entire agent tree with a single signal.

Sema Code supports one-level task delegation through a dedicated agent-spawning mechanism, where the main agent can create sub-agents but sub-agents cannot spawn further sub-agents---bounding call depth to avoid unbounded context and resource consumption. Each sub-agent $a_i$ owns a fully isolated state space $\mathcal{S}_{a_i} = \langle H_{a_i},\ T_{a_i},\ F_{a_i},\ S_{a_i} \rangle$; the only shared resource with the main agent is the abort controller $\mathcal{C}_{\text{abort}}$. Sub-agent lifecycle follows a strict three-phase protocol:
\begin{enumerate}
\item \textbf{Creation:} The engine assigns a unique session identifier, constructs a dedicated system prompt encompassing environment rules and repository status, and provisions a restricted toolset that explicitly omits the delegation capability to bound call depth.
\item \textbf{Execution:} The sub-agent evaluates its reasoning loop entirely within its private state boundary, ensuring that verbose tool outputs and intermediate message histories do not pollute the main agent's context.
\item \textbf{Cleanup:} The system atomically purges the localized state to prevent memory leaks, aggregates telemetry data such as token consumption and execution time, and surfaces only the final synthesized result back to the delegating parent.
\end{enumerate}

While multi-agent scheduling handles task decomposition, it introduces a new challenge: when multiple agents run concurrently or sequentially, the user must be able to interrupt the entire agent tree at any point. The next subsection describes how interrupt signals propagate reliably across the agent hierarchy.

\subsection{Interrupt Propagation and Tool Scheduling}
\label{sec:interrupt}

To handle asynchronous interrupts arriving at any point in the query lifecycle, the runtime enforces strict boundary controls across four distinct execution phases:
\begin{itemize}
\item \textbf{Post-Inference Dispatch:} If an interrupt occurs after the LLM generates a plan but before execution begins, the system generates cancellation placeholders for all pending operations to prevent cascading downstream API errors.
\item \textbf{Pre-Execution Verification:} If interrupted immediately before an individual tool is invoked, the engine bypasses the operation entirely and yields a cancellation context.
\item \textbf{Active Execution Monitoring:} During an ongoing tool invocation, the engine distinguishes between user-initiated refusals (which are permanently retained as contextual history) and standard abort signals.
\item \textbf{Recursion Termination:} After all scheduled tools conclude and before the next recursive reasoning step, the runtime flushes unfinished results and securely collapses the recursion boundary.
\end{itemize}

Beyond interrupt handling, tool execution order is governed by a strict read-write scheduling policy. When a scheduled batch consists entirely of read-only operations (e.g., file and content searches), the engine exploits I/O concurrency to execute them in parallel. However, if the batch contains any write operation (e.g., file edits or shell executions), the engine strictly serializes the entire execution to prevent race conditions. This reasoning-acting alternation inherits the core philosophy of ReAct~\cite{yao2023react} while introducing read-write-aware concurrency at the scheduling layer.

\subsection{Intelligent Todo-based Process Managemen}
\label{sec:todo}

Task decomposition and interrupt handling ensure correct execution, but a critical usability challenge remains: maintaining observability. During complex tasks spanning dozens of tool invocations, users struggle to track progress without structured feedback. Furthermore, because LLM text generation is non-deterministic, successive updates to a task list often introduce slight phrasing variations (e.g., altering Refactoring auth module'' to Refactor authentication module''). Naively rendering these updates causes disruptive UI flickering, as items appear to change even when only their underlying status has advanced.

To resolve this, Sema Core implements a deterministic task-tracking state machine. Each task item is tracked via a unique identifier, its descriptive content, and a discrete lifecycle state (such as pending, active, or completed). At the input validation layer, the engine enforces a strict mutual-exclusion constraint: at most one subtask can be marked as actively in progress at any given time. This guarantees that users always possess a clear understanding of the agent's immediate focus.

The state update algorithm dynamically distinguishes between partial status transitions and full list replacements. When a newly emitted task list contains only previously registered identifiers, the engine performs a subset update, modifying only the lifecycle states while intentionally preserving the original descriptive text. This completely eliminates the UI flicker caused by LLM rephrasing. Conversely, the introduction of unknown identifiers triggers a full replacement to register new subtasks. Finally, to maintain a clean global context, task updates from sub-agents are strictly scoped to their local lifecycles and do not pollute the main agent's presentation layer.

\subsection{Background Task Execution}
\label{sec:bgtask}

While the preceding subsystems orchestrate multi-agent delegation, a critical bottleneck remains at the execution layer: shell operations frequently involve long-running processes (e.g., compilations or test suites) that block the agent's synchronous reasoning loop. Naive timeout mechanisms are inadequate, as they destructively terminate processes, discard partial outputs, and fail to release shared process resources (e.g., the primary shell singleton), thereby deadlocking subsequent operations.

Sema Core resolves this through an architectural separation of execution and observation. Long-running tasks are offloaded to a dedicated background manager, allowing the primary dialog loop to resume user interaction immediately. Each background task follows a strict, deterministic lifecycle, transitioning from an active running state to either a successful completion, a failure (based on process exit codes), or a forcefully stopped state. To ensure resource stability, the runtime strictly bounds both the maximum concurrent executions and the retention pool of historical task states to prevent memory leaks.

\textbf{Triggering Mechanisms.} Background execution is invoked via two distinct routing paths. The \textit{proactive route} occurs when the agent explicitly flags an operation for asynchronous processing; the engine immediately spawns an isolated system process and returns a tracking identifier. More critically, the \textit{reactive takeover route} gracefully handles operations that unexpectedly exceed execution thresholds. Instead of terminating the timed-out process, the engine dynamically detaches the active shell session and its associated temporary file descriptors, handing them over entirely to the background manager. The system then cleanly resets the primary shell singleton. This robust decoupling perfectly resolves the tension between maintaining an unblocked conversational interface and preserving the continuity of heavy workloads.

\textbf{Output Management and Asynchronous Callbacks.} To manage task outputs, the system employs a dual-write strategy across memory and disk, ensuring both low-latency streaming and persistent snapshotting. The output observer utilizes an adaptive polling mechanism that progressively reduces its sampling frequency during extended executions to conserve CPU cycles. The agent interacts with these background processes through a streamlined abstraction, allowing it to trigger tasks, query real-time snapshots, or issue termination signals. Upon task completion, the engine injects a structured notification directly into the main agent's context. This asynchronous callback acts as a wake-up signal, prompting the LLM to analyze the final results and effectively closing the loop between background execution and active reasoning.

\section{Security and Ecosystem}

Consider a prompt-injection attack in which a malicious repository's \texttt{README} instructs the coding agent to execute \texttt{rm -rf} or, more subtly, to exfiltrate environment variables through a crafted \texttt{curl} command embedded in a build script. When the agent operates with unrestricted shell access, a single compromised instruction can destroy a production file system or leak API keys. Recent analyses confirm that LLM-based agents inherit and amplify such risks because they autonomously compose tool invocations that no human reviews in real time~\cite{ruan2024lmagent_risks,siu2026llmagent_security}. Yet overly restrictive guardrails cripple productivity: an agent that demands approval for every read-only \texttt{git status} becomes an obstacle rather than an accelerator. Sema Code addresses this tension through a layered permission architecture that confines risk without impeding routine work, paired with an open ecosystem that extends agent capabilities through governed, discoverable channels.

\subsection{Four-Layer Permission Decision}

NeMo Guardrails~\cite{rebedea2023nemo_guardrails} provides programmable safety rails at the conversational layer, but AI coding requires finer-grained control at the execution layer, where the agent directly manipulates files, shells, and external services.

The four layers correspond to the four fundamental operation types through which an agent interacts with its environment: modifying project artifacts (file edits), executing system commands (shell), invoking internal capabilities (Skills), and calling external services (MCP). Each type carries a distinct risk profile: file edits are reversible within a session, shell commands can have irreversible side effects, Skill invocations reshape agent behavior, and MCP calls cross trust boundaries into third-party systems. Consequently, this variance motivates separate permission semantics rather than a single allow/deny gate.

Formally, the system implements a decision function $P: \mathcal{O} \times \mathcal{C} \rightarrow \{\text{allow},\ \text{deny},\ \text{request}\}$, where an operation $\mathcal{O}$ evaluated under the current session context $\mathcal{C}$ yields one of three terminal states. To accommodate the distinct risk profiles discussed above, $P$ routes the incoming request to one of four specialized evaluation layers, as detailed in Table~\ref{tab:permission}.

\begin{table}[t]
\centering
\small
\begin{tabular}{@{}lllll@{}}
\toprule
Layer & Operation Type & Fast-Pass & Approval Trigger & Granularity \\
\midrule
L1 & File editing & Edit flag on & First edit request & Session \\
L2 & Shell commands & Prefix in whitelist & Not whitelisted / injection & Project \\
L3 & Skill invocation & Skill authorized & First invocation & Project \\
L4 & MCP operations & Name authorized & First invocation & Project \\
\bottomrule
\end{tabular}
\caption{Four-layer permission decision matrix.}
\label{tab:permission}
\end{table}

Among these layers, shell execution (L2) presents the most acute security challenge due to the pervasive risk of command injection. The shell command layer handles pipeline composition and injection detection via a two-stage evaluation. The first stage is a deterministic whitelist check:
\begin{equation}
P_{\text{Bash}}(c) = \text{allow}
\iff
\begin{cases}
\text{head}(c) \in \mathcal{W},
& \text{if } c \text{ is a simple command}, \\[4pt]
\forall i \in \{1,\dots,p\},\ \text{head}(c_i) \in \mathcal{W},
& \text{if } c = c_1 \circ c_2 \circ \dots \circ c_p.
\end{cases}
\end{equation}
where $\circ \in \{\texttt{|},\ \texttt{\&\&},\ \texttt{;}\}$ represents any shell composition operator. Each sub-command head is independently matched against the predefined whitelist $\mathcal{W}$.

If any sub-command falls outside the whitelist, the operation defaults to the \texttt{request} state. However, to prevent users from blindly approving obfuscated malicious payloads, the system triggers a secondary LLM-assisted static analysis before surfacing the prompt. The LLM decomposes the command string into sub-command prefixes and classifies each against known injection patterns (e.g., backtick substitution, process substitution via \texttt{\$(...)}, and chained destructive operations). If a payload is identified as an injection risk, the system either escalates the request with a severe security warning or automatically downgrades to \texttt{deny}. This LLM-assisted detection effectively catches adversarial obfuscation that static regex matching would miss, though we acknowledge it inherits the base model's false-negative rate and incurs one inference round-trip of latency.



\subsection{Asynchronous Approval Protocol}

To prevent permission gates from blocking the core reasoning loop, Sema Core implements an event-driven, asynchronous approval protocol. When an operation requires user consent, the engine emits an authorization request to the event bus and strictly suspends the active execution context. This decoupling of the engine's state machine from the presentation layer allows arbitrary clients (e.g., IDE dialogs, Web UIs, messaging bots) to handle the user interaction natively. The engine remains paused until it receives a concrete user decision or a global abort signal. Upon receiving the response, the protocol resumes execution via one of four distinct resolution paths: transient approval (allowing the single operation), persistent authorization (updating the project-level security policy), explicit rejection (triggering an immediate abort), or user-guided correction (injecting natural language feedback directly into the tool result for the agent to reconsider). Furthermore, to minimize prompt fatigue during complex delegations, sub-agents implicitly inherit the permission boundaries of their parent, ensuring that structural security does not degrade multi-agent efficiency.



\subsection{Three-Tier Marketplace Ecosystem}

Sema Code's ecosystem strategy rests on two premises: the MCP protocol~\cite{anthropic2024mcp} and Skill/Plugin specifications are de facto standards that frameworks should support natively rather than redefine~\cite{hou2025mcp_survey}; and ecosystem value lies in discoverability---one-click browsing and installation, not manual configuration editing.

A unified plugin system would conflate three fundamentally different granularities of capability extension. \textbf{MCP services} operate at the \emph{infrastructure} granularity: each service wraps an external system (database, browser, API gateway) behind a standardized protocol, exposing coarse-grained tools that the agent invokes as black-box operations. \textbf{Skills} operate at the \emph{behavior} granularity: a Skill is a Markdown document with structured metadata that reshapes \emph{how} the agent reasons---its prompting strategy, constraints, and model preferences---without introducing new tool endpoints. \textbf{Plugins} operate at the \emph{workflow} granularity: they hook into the engine's command and lifecycle system to orchestrate multi-step processes (code review pipelines, deployment gates) that span multiple tool invocations. Collapsing these three layers into one would either over-privilege lightweight behavior modifiers with tool-level access or under-serve infrastructure integrations that require persistent connections and authentication management.

\textbf{MCP Service Marketplace.} Users browse hundreds of pre-integrated MCP services (databases, browser automation, code analysis, API gateways) from the ClaWHub marketplace (the project's online extension registry), previewing capabilities and installing to global or project configuration with one click. Installed services auto-load at session startup, dynamically injecting capabilities into the LLM's available set.

\textbf{Skill Marketplace.} Skills are Markdown capability files with YAML frontmatter declaring scenarios, constraints, and model preferences. The marketplace curates collections for code review, test generation, documentation, and security analysis. One-click installation via CLI or VSCode registers Skills to project configuration; the LLM loads available Skills by priority (project $\succ$ user $\succ$ built-in) and invokes as needed.

\textbf{Plugin Ecosystem (Command / Hook).} Command plugins extend the slash command system (e.g., \texttt{/review}, \texttt{/deploy}). Hook plugins register lifecycle callbacks around operations for logging, compliance auditing, and custom validation. All plugins are distributed through ClaWHub, supporting enable/disable/update without session restart.

\subsection{Multi-Model Adaptation Layer}

Sema Core ships two built-in adapters---Anthropic native and OpenAI-compatible---exposing a unified streaming interface that normalizes provider-specific features (Extended Thinking, Prompt Caching, tool-use schemas) into a common event protocol. This normalization is non-trivial: Anthropic's streaming API emits thinking blocks as first-class events, whereas OpenAI-compatible endpoints encode reasoning tokens opaquely; Sema Core's adapter layer reconciles these differences so that upper-layer logic (permission checks, context compression, sub-agent delegation) operates identically regardless of the underlying provider. RouteLLM~\cite{ong2025routellm} has shown that preference-based model routing can halve costs without sacrificing quality. Via \texttt{customBaseUrl}, Sema Core connects seamlessly to open-source models (Code Llama~\cite{roziere2023codellama}, DeepSeek-Coder~\cite{guo2024deepseek_coder}) and commercial models (DeepSeek V3/R1, GLM-5, Qwen3-235B, Gemini 2.5 Pro). The same agent capabilities run uniformly across different models, with upper-layer applications unaware of underlying differences.

\section{Deployment Validation}
\label{sec:validation}

To validate Sema Code's core architectural claim---that a single, decoupled engine can power fundamentally different product forms without per-product modification---we report on two deployments built on the same Sema Core engine: a VSCode extension for individual developers and SemaClaw, a multi-channel agent platform for team collaboration. Rather than benchmarking individual mechanisms in isolation, we focus on the architectural question that motivates this work: can the same engine, without any modification, serve clients with fundamentally different interaction models, concurrency requirements, and deployment topologies?

\subsection{IDE Integration: VSCode Extension}
\label{sec:vscode}

The Sema Code Extension imports Sema Core directly via npm, running within the VSCode Extension Host process and eliminating CLI subprocess overhead. It subscribes to Sema Core's event stream for UI rendering, providing real-time diff preview with one-click rollback, a Todo task board, background task panel, session history management, and runtime model hot-switching. Published on the Open VSX marketplace for Windows, macOS, and Linux.

In this deployment, the extension operates in single-user, in-process mode. It exercises the following engine mechanisms: adaptive context compression during extended refactoring sessions, the four-layer permission system via native VSCode dialogs, background task execution for long-running builds and tests, and the MCP/Skill/Plugin ecosystem through the integrated marketplace. The single-user deployment mode means that multi-tenant isolation is not exercised, making it a complementary case study to SemaClaw.

\subsection{Multi-Channel Agent Platform: SemaClaw}
\label{sec:semaclaw}

SemaClaw is a multi-channel AI agent platform built on Sema Core, importing the same npm package version to power core capabilities while adding Telegram, Feishu, and other messaging channel access with a Web UI management interface. Unlike the VSCode extension, SemaClaw runs as a server-side Node.js process serving multiple concurrent users.

SemaClaw exercises engine mechanisms that the VSCode extension does not: multi-tenant isolation with concurrent users sharing a single process, the FIFO input queue absorbing message bursts from messaging channels, and the asynchronous approval protocol adapting permission UI to inline messaging buttons rather than modal dialogs. Together, the two deployments cover all eight mechanisms.

\subsection{Architectural Observations}
\label{sec:arch_observations}

Three observations from deploying the same engine across these two products are worth highlighting.

\textbf{Zero engine modifications.} Neither product required any change to the Sema Core codebase. All behavioral differences---UI rendering, channel routing, permission dialogs, deployment topology---are realized entirely at the client layer. This confirms that the three-layer separation achieves its design goal: the engine is truly delivery-agnostic.

\textbf{Complementary mechanism coverage.} The two deployments exercise different subsets of the engine's mechanisms. The VSCode extension stresses context compression and background task execution in long single-user sessions; SemaClaw stresses multi-tenant isolation and input queuing under concurrent multi-user load. The fact that both products run correctly on the same engine binary, without feature flags or conditional paths, provides evidence that the mechanisms compose without interference.

\textbf{Client-side integration effort.} Building the VSCode extension required implementing event stream rendering and VSCode-specific UI components; building SemaClaw required implementing channel adapters (Telegram Bot API, Feishu webhook) and a multi-session manager. In both cases, the client developer interacted exclusively with Sema Core's public API and the typed event stream, without needing to understand the engine's internal state management, compression logic, or permission machinery. This clean separation of concerns validates the architectural efficacy of the embeddable engine.

\subsection{Limitations}
\label{sec:limitations}

This work has several limitations. First, the deployment validation covers only two product forms; whether the architecture generalizes to other form factors (e.g., CI/CD pipeline integration, Jupyter notebook embedding) remains to be demonstrated. Second, the current deployments have not undergone stress testing at scale---production workloads with hundreds of concurrent users may reveal performance bottlenecks not observed in our setting. Third, context compression quality depends on the underlying LLM's summarization ability; highly technical conversations with dense code fragments may lose critical details during compression, and the degradation path (truncation) sacrifices coherence for availability. Fourth, the cross-language integration interfaces (WebSocket, gRPC), while functionally complete, have not been benchmarked for streaming throughput or error recovery under adversarial network conditions. Finally, all current deployments use a single Sema Core process; horizontal scaling across multiple engine instances introduces state synchronization challenges that the current architecture does not address.

\section{Conclusion}

We have presented Sema Code, an open framework that elevates AI coding capabilities from product-locked features to programmable infrastructure. By fully decoupling the core engine from all client layers, Sema Code makes embedding an AI coding agent as simple as adding an npm dependency. Eight mechanisms---from multi-tenant isolation to background task execution---systematically address the engineering challenges of exposing an agent engine as a public API.

We argue that the next critical evolution for AI coding frameworks is embeddability and open ecosystems. When coding capabilities are no longer locked in specific products but integrated as engines into existing toolchains, and when capabilities accumulate through open marketplaces, the developer ecosystem will see deeper transformation. Sema Code is our systematic exploration in this direction.

Future work will pursue three directions: vector-retrieval-based context management for finer-grained historical recall beyond compression; extended multi-agent coordination supporting inter-agent state sharing and message passing, evolving from delegation to cooperation; and distributed task scheduling across engine instances for elastic production deployment.

\clearpage
\newpage
\bibliographystyle{assets/plainnat}
\bibliography{paper}

@article{chen2021codex,
  title={Evaluating Large Language Models Trained on Code},
  author={Chen, Mark and Tworek, Jerry and Jun, Heewoo and Yuan, Qiming and Pinto, Henrique Ponde de Oliveira and Kaplan, Jared and Edwards, Harri and Burda, Yuri and Joseph, Nicholas and Brockman, Greg and others},
  journal={arXiv preprint arXiv:2107.03374},
  year={2021}
}

@article{peng2023copilot_impact,
  title={The Impact of {AI} on Developer Productivity: Evidence from {GitHub Copilot}},
  author={Peng, Sida and Kalliamvakou, Eirini and Cihon, Peter and Demirer, Mert},
  journal={arXiv preprint arXiv:2302.06590},
  year={2023}
}

@misc{cursor2024,
  title={The {AI} Code Editor},
  author={{Cursor}},
  howpublished={\url{https://cursor.com}},
  year={2024}
}

@techreport{anthropic2025claude_code,
  title={Claude Code: Agentic Coding in the Terminal},
  author={{Anthropic}},
  institution={Anthropic},
  year={2025}
}

@inproceedings{yang2024sweagent,
  title={{SWE}-agent: Agent-Computer Interfaces Enable Automated Software Engineering},
  author={Yang, John and Jimenez, Carlos E. and Wettig, Alexander and Lieret, Kilian and Yao, Shunyu and Narasimhan, Karthik and Press, Ofir},
  booktitle={Advances in Neural Information Processing Systems (NeurIPS)},
  year={2024}
}

@article{xia2024agentless,
  title={Agentless: Demystifying {LLM}-based Software Engineering Agents},
  author={Xia, Chunqiu Steven and Deng, Yinlin and Dunn, Soren and Zhang, Lingming},
  journal={arXiv preprint arXiv:2407.01489},
  year={2024}
}

@inproceedings{yang2024swebench,
  title={{SWE}-bench: Can Language Models Resolve Real-World {GitHub} Issues?},
  author={Yang, John and Jimenez, Carlos E. and Wettig, Alexander and Yao, Shunyu and Pei, Kexin and Press, Ofir and Narasimhan, Karthik},
  booktitle={International Conference on Learning Representations (ICLR)},
  year={2024}
}

@article{hou2024llm4se_survey,
  title={Large Language Models for Software Engineering: A Systematic Literature Review},
  author={Hou, Xinyi and Zhao, Yanjie and Liu, Yue and Yang, Zhou and Wang, Kailong and Li, Li and Luo, Xiapu and Lo, David and Grundy, John and Wang, Haoyu},
  journal={ACM Transactions on Software Engineering and Methodology (TOSEM)},
  year={2024}
}

@article{gaffney2022sqlite,
  title={{SQLite}: Past, Present, and Future},
  author={Gaffney, Kevin P. and Prammer, Martin and Brasfield, Larry and Hipp, D. Richard and Kennedy, Dan and Patel, Jignesh M.},
  journal={Proceedings of the VLDB Endowment},
  volume={15},
  number={12},
  year={2022}
}

@misc{github2023copilot,
  title={{GitHub Copilot}: Your {AI} pair programmer},
  author={{GitHub}},
  year={2023}
}

@article{wang2024openhands,
  title={{OpenHands}: An Open Platform for {AI} Software Developers as Generalist Agents},
  author={Wang, Xingyao and others},
  journal={arXiv preprint arXiv:2407.16741},
  year={2024}
}

@inproceedings{hong2024metagpt,
  title={{MetaGPT}: Meta Programming for A Multi-Agent Collaborative Framework},
  author={Hong, Sirui and Zhuge, Mingchen and Chen, Jonathan and others},
  booktitle={International Conference on Learning Representations (ICLR)},
  year={2024}
}

@inproceedings{qian2024chatdev,
  title={{ChatDev}: Communicative Agents for Software Development},
  author={Qian, Chen and Liu, Wei and Liu, Hongzhang and others},
  booktitle={Annual Meeting of the Association for Computational Linguistics (ACL)},
  year={2024}
}

@article{wu2023autogen,
  title={{AutoGen}: Enabling Next-Gen {LLM} Applications via Multi-Agent Conversation},
  author={Wu, Qingyun and Bansal, Gagan and Zhang, Jieyu and others},
  journal={arXiv preprint arXiv:2308.08155},
  year={2023}
}

@misc{chase2022langchain,
  title={{LangChain}: Building applications with {LLMs} through composability},
  author={Chase, Harrison},
  howpublished={GitHub},
  year={2022}
}

@inproceedings{wei2022cot,
  title={Chain-of-Thought Prompting Elicits Reasoning in Large Language Models},
  author={Wei, Jason and Wang, Xuezhi and Schuurmans, Dale and others},
  booktitle={Advances in Neural Information Processing Systems (NeurIPS)},
  year={2022}
}

@inproceedings{yao2023react,
  title={{ReAct}: Synergizing Reasoning and Acting in Language Models},
  author={Yao, Shunyu and Zhao, Jeffrey and Yu, Dian and Du, Nan and Shafran, Izhak and Narasimhan, Karthik and Cao, Yuan},
  booktitle={International Conference on Learning Representations (ICLR)},
  year={2023}
}

@inproceedings{schick2023toolformer,
  title={Toolformer: Language Models Can Teach Themselves to Use Tools},
  author={Schick, Timo and Dwivedi-Yu, Jane and Dess{\`i}, Roberto and others},
  booktitle={Advances in Neural Information Processing Systems (NeurIPS)},
  year={2023}
}

@inproceedings{patil2024gorilla,
  title={Gorilla: Large Language Model Connected with Massive {APIs}},
  author={Patil, Shishir G. and Zhang, Tianjun and Wang, Xin and Gonzalez, Joseph E.},
  booktitle={Advances in Neural Information Processing Systems (NeurIPS)},
  year={2024}
}

@inproceedings{qin2024toolllm,
  title={{ToolLLM}: Facilitating Large Language Models to Master 16000+ Real-world {APIs}},
  author={Qin, Yujia and Liang, Shihao and Ye, Yining and others},
  booktitle={International Conference on Learning Representations (ICLR)},
  year={2024}
}

@article{packer2023memgpt,
  title={{MemGPT}: Towards {LLMs} as Operating Systems},
  author={Packer, Charles and Wooders, Sarah and Lin, Kevin and Fang, Vivian and Patil, Shishir G. and Stoica, Ion and Gonzalez, Joseph E.},
  journal={arXiv preprint arXiv:2310.08560},
  year={2023}
}

@inproceedings{jiang2023llmlingua,
  title={{LLMLingua}: Compressing Prompts for Accelerated Inference of Large Language Models},
  author={Jiang, Huiqiang and Wu, Qianhui and Lin, Chin-Yew and Yang, Yuqing and Qiu, Lili},
  booktitle={Conference on Empirical Methods in Natural Language Processing (EMNLP)},
  year={2023}
}

@inproceedings{li2025prompt_compression_survey,
  title={Prompt Compression for Large Language Models: A Survey},
  author={Li, Zongqian and Liu, Yinhong and Su, Yixuan and Collier, Nigel},
  booktitle={Annual Conference of the North American Chapter of the Association for Computational Linguistics (NAACL)},
  year={2025}
}

@article{wang2024agent_survey,
  title={A Survey on Large Language Model based Autonomous Agents},
  author={Wang, Lei and Ma, Chen and Feng, Xueyang and others},
  journal={Frontiers of Computer Science},
  year={2024}
}

@inproceedings{barros2022lsp,
  title={Editing Support for Software Languages: Implementation Practices in Language Server Protocols},
  author={Barros, Djamel Eddine Khelladi and Peldszus, Sven and Assuncao, Wesley K. G. and Berger, Thorsten},
  booktitle={ACM/IEEE International Conference on Model Driven Engineering Languages and Systems (MODELS)},
  year={2022}
}

@inproceedings{peng2024yarn,
  title={{YaRN}: Efficient Context Window Extension of Large Language Models},
  author={Peng, Bowen and Quesnelle, Jeffrey and Fan, Honglu and Shippole, Enrico},
  booktitle={International Conference on Learning Representations (ICLR)},
  year={2024}
}

@inproceedings{ruan2024lmagent_risks,
  title={Identifying the Risks of {LM} Agents with an {LM}-Emulated Sandbox},
  author={Ruan, Yangjun and Dong, Honghua and Wang, Andrew and others},
  booktitle={International Conference on Learning Representations (ICLR)},
  year={2024}
}

@article{siu2026llmagent_security,
    title={A Framework for Formalizing {LLM} Agent Security},
    author={Siu, Vincent and He, Jingxuan and Montgomery, Kyle and Wang, Zhun and Gong, Neil and Wang,Chenguang and Song, Dawn},
    journal={arXiv preprint arXiv:2603.19469},
    year={2026}
}

@inproceedings{rebedea2023nemo_guardrails,
  title={{NeMo Guardrails}: A Toolkit for Controllable and Safe {LLM} Applications with Programmable Rails},
  author={Rebedea, Traian and Dinu, Razvan and Sreedhar, Makesh Narsimhan and Parisien, Christopher and Cohen, Jonathan},
  booktitle={Conference on Empirical Methods in Natural Language Processing: System Demonstrations (EMNLP)},
  year={2023}
}

@misc{anthropic2024mcp,
  title={Model Context Protocol Specification},
  author={{Anthropic}},
  howpublished={\url{https://modelcontextprotocol.io}},
  year={2024}
}

@article{hou2025mcp_survey,
  title={Model Context Protocol ({MCP}): Landscape, Security Threats, and Future Research Directions},
  author={Hou, Xinyi and Zhao, Yanjie and Wang, Shenao and Wang, Haoyu},
  journal={ACM Transactions on Software Engineering and Methodology (TOSEM)},
  year={2025}
}

@inproceedings{ong2025routellm,
  title={{RouteLLM}: Learning to Route {LLMs} from Preference Data},
  author={Ong, Isaac and Almahairi, Amjad and Wu, Vincent and others},
  booktitle={International Conference on Learning Representations (ICLR)},
  year={2025}
}

@article{roziere2023codellama,
  title={Code Llama: Open Foundation Models for Code},
  author={Rozi{\`e}re, Baptiste and Gehring, Jonas and Gloeckle, Fabian and others},
  journal={arXiv preprint arXiv:2308.12950},
  year={2023}
}

@article{guo2024deepseek_coder,
  title={{DeepSeek-Coder}: When the Large Language Model Meets Programming -- The Rise of Code Intelligence},
  author={Guo, Daya and Zhu, Qihao and Yang, Dejian and others},
  journal={arXiv preprint arXiv:2401.14196},
  year={2024}
}


\end{document}